# Some computational aspects of using Huygens-Fresnel-Kirchoff diffraction theory

Ilya A. Kudryavtsev

Samara National Research University, Moskovskoye shosse, 34, Samara, Russia

rtf@ssau.ru

Diffraction is a phenomenon, discussed for centuries from various points of view. The very simple principle, proposed by Huygens [1] and then modified by Fresnel[2], Stokes [3] and Kirchoff [4], allows us to make calculations, substituting an incident wave by the multitude of waves, radiated by the number of secondary sources with regard for interference. Besides mentioned scientists, many others have contributed to this theory. A reasonable historical review can reveal a challenging history of the research of this phenomenon with many names and this story is not still ended. A good historical review can be seen in [5]. Classical cases of the diffraction by a hole and a disk are included in many textbooks for scholars. Nevertheless, there is still some place for discussions. This paper is an attempt to discuss some computational issues, based on the original Huygens-Fresnel-Kirchoff method and apply it to 3D objects.

1. General considerations

The idea, proposed by Huygens [1] and then enriched by Fresnel [2], is based on the substitution of an incident wave by the result of superposition of secondary spherical waves, radiated by a number of sources, excited by the incident wave. The reason, why it is possible, was explained by the oscillations of ether molecules, excited by an incident wave. Huygens used this approach to explain reflection and refraction. A.Fresnel suspected that these secondary sources were not isotropic, however avoided this topic with elegance in his reconstruction [2]. Exactly his words were: «*La recherche de la loi suivant laquelle leur intensité varierait autour de chaque centre d'ébranlement, présenterait sans doute de grandes dificultés: mais heureusement nous n'avons pas besoin de la connaître;*».

Next, Stokes [3] using the same ideas about elasticity of the ether, came to an idea of the nonisotropic radiation of secondary waves, but discussed also an idea of isotropic radiation. It is important to add that he noted, "*…in passing from the primary to secondary waves it is necessary to accelerate the waves by a quarter of undulation*". This inference was also drawn by A.Smith [6], referred by Stokes.

G. Kirchoff [4] used an approach that is more complicated (and more rigorous), based on Helmholtz equation, however, the simplicity was lost. With some limitations and assumptions, the following formula for the case, shown in fig.1, was derived [7]:

$$U(P) = -\frac{iA}{2\lambda} \iint_A \frac{\exp[ik(r+s)]}{rs} \cdot [\cos(n,r) - \cos(n,s)] dS$$

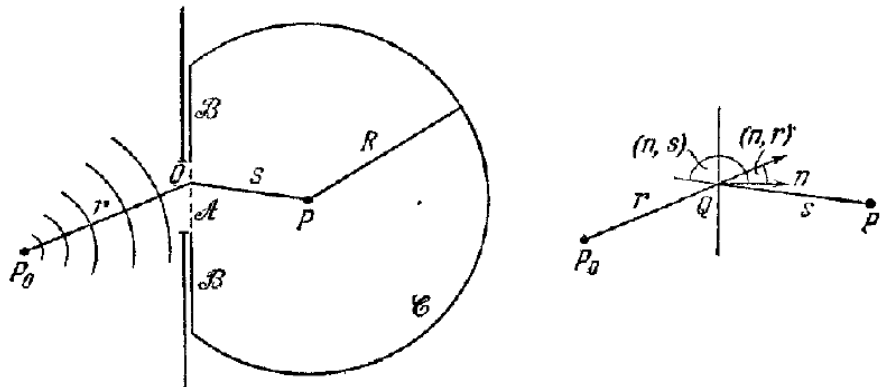

Fig. 1. Schematic diffraction setup [6].

When the incident wave is spherical and we use its wavefront as a surface where secondary sources are located, Kirchoff's approach provides an obliquity factor:

$$K(\chi) = -\frac{i}{2\lambda}(1 + \cos(\chi))$$

where $\chi = \pi - (n,s)$

The same result was obtained by Stokes. In [8] a reader can find a discussion about various ways of calculations, including various versions of obliquity factor, called also incident factor.

Summarizing the ideas of classics, it is possible to offer a computational algorithm, based on the following issues:

1. An incident wave of the frequency $\omega$ with the intensity $I_i$ can be substituted by the set of secondary radiators dS, which are excited by the incident wave. Every such source transmits a secondary spherical wave, however with some directivity factor $K_0$, imposed by the obliquity factor. The intensity of this wave of the magnitude $I_{im}$ can be expressed as
$$I_i = I_{im} \cdot \cos(\omega t - kz) \text{ in case of a plane wave}$$
$$I_i = \frac{I_{im}}{z} \cdot \cos(\omega t - kz) \text{ in case of a spherical wave}$$

2. The magnitude of a secondary wave $I_{sm}$ as shown in fig.2 can be expressed as:
$$I_{sm} = I_{im}(dS) \cdot dS \cdot K_0$$
where $I_{im}(dS)$ is an intensity of the incident wave at dS.

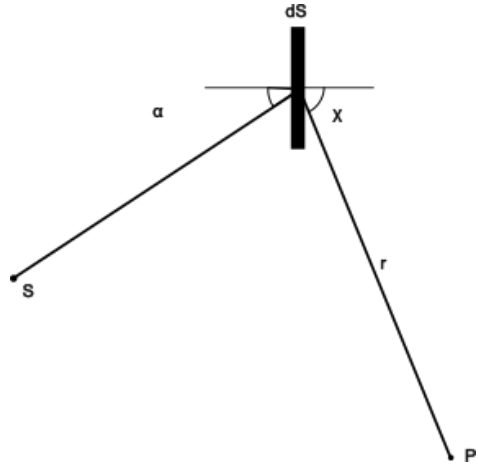

Fig. 2. To the simulation algorithm.

3. The disturbance in an arbitrary point P can be calculated as a result of the interference of secondary waves:
$$U(r, \omega t) = \sum \frac{I_{im}(dS)}{\lambda r} \cdot dS \cdot K_0 \cdot \cos(\omega t - kr - \frac{\pi}{2})$$
Here $r$ is a distance from a secondary source to P, $\lambda$- wavelength, $k = \frac{2\pi}{\lambda}$. The summation must be performed with regard for the direction of the propagation of a secondary wave.

4. The magnitude of the resulting wave can be calculated, using an auxiliary secondary wave, shifted in phase by $\frac{\pi}{2}$ with the following expression:
$$U_m = \sqrt{U(r, \omega t)^2 + U(r, \omega t - \frac{\pi}{2})^2}$$

5. In case, when a diffraction pattern is needed, the resulting disturbance can be calculated, using the Babinet's principle in a following form:
$$D(r, \omega t) = I_i(r, \omega t) - U(r, \omega t)$$
where $I_i(r, \omega t)$ is an intensity of the primary wave at point P.

In this paper, five cases of $K_0$ are discussed and compared:
1. Isotropic secondary source ($K_0 = 1$)
2. Isotropic secondary source, excited by the primary wave with regard for $\cos(\alpha)$ ($K_0 = \cos(\alpha)$)
3. Non-isotropic secondary source ($K_0 = \cos(\chi)$), regardless of $\alpha$
4. Non-isotropic secondary source ($K_0 = \cos(\chi) \cdot \cos(\alpha)$) with regard for $\alpha$
5. Kirchoff-Stokes secondary source ($K_0 = \frac{\cos(\alpha)+\cos(\chi)}{2}$)

## 2. Wave reconstruction

To be sure in correctness of the computational algorithm, it is reasonable to verify its accuracy, comparing parameters of the incident wave and the result of the interference of secondary waves in case, when no obstruction is present. In case of a plane wave, for example, its magnitude must be the same at any point. Since we discuss a numerical approach, it is essential to verify the accuracy with some reasonable approximation. Here and further 1% deviation threshold is used. The accuracy can be improved by increasing the number of secondary sources uses, less dS values etc.

In case of plane waves it is essential to use the wavefront as a place, where the secondary sources are located. Obviously, $\cos(\alpha) = 1$, therefore only cases 1, 4, 5 are to be investigated. Fig.3 shows the geometry of the simulation. Magnitudes of the disturbance along both axis Ox and Oz must be uniform, waveform of the simulated wave must coincide with the incident wave as well. Obviously, components Ux, received from all sources will compensate each other, when Uz will form resulting intensity. In the experiment $dS = dr \cdot dr$, where $dr = \lambda/10$ (arbitrarily).

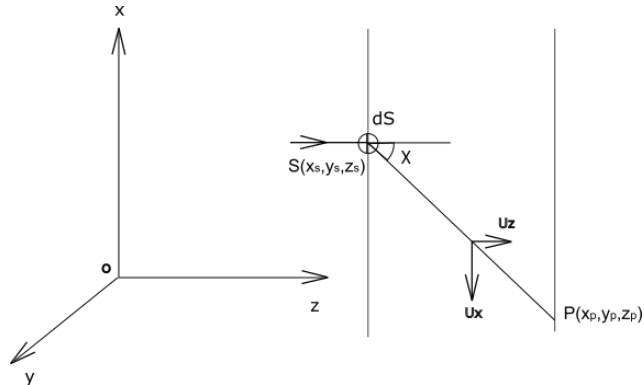

Fig. 3. Simulation of the plane wave.

In the experiment the intensity of the primary wave ($I_i$) was set to 1. Z = 0 corresponds to the plane (wavefront), where secondary sources are located. In fig. 4 one can see the results of simulation in case 1 ($K_0 = 1$). Here and further $10^8$ secondary sources were used.

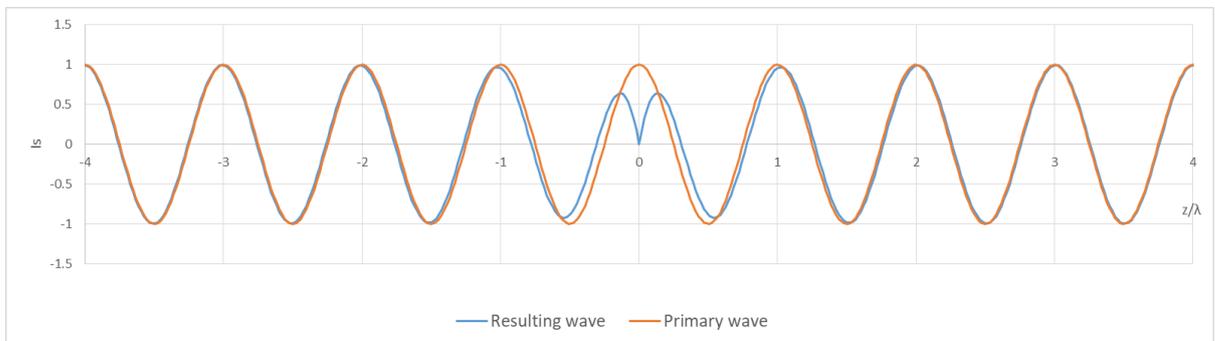

Fig. 4. Wave intensity along Oz axis.

It can be seen that the intensity of the primary wave and the result of interference of secondary waves ($I_s$ - resulting wave) coincide closely at all points except the area near $z = 0$. This behavior can be explained by the phase shift of secondary waves.

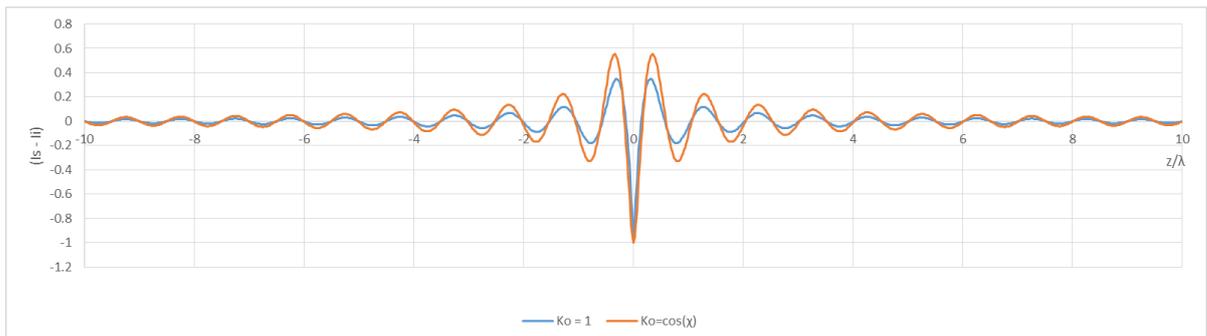

Fig. 5. Comparison of the discrepancy ($I_s - I_i$) between the primary and resulting waves.

In Fig.5 it could be seen that with $K_0 = 1$ (case 1) resulting wave convergence is better, however the difference becomes negligible, when the distance from secondary sources increases.

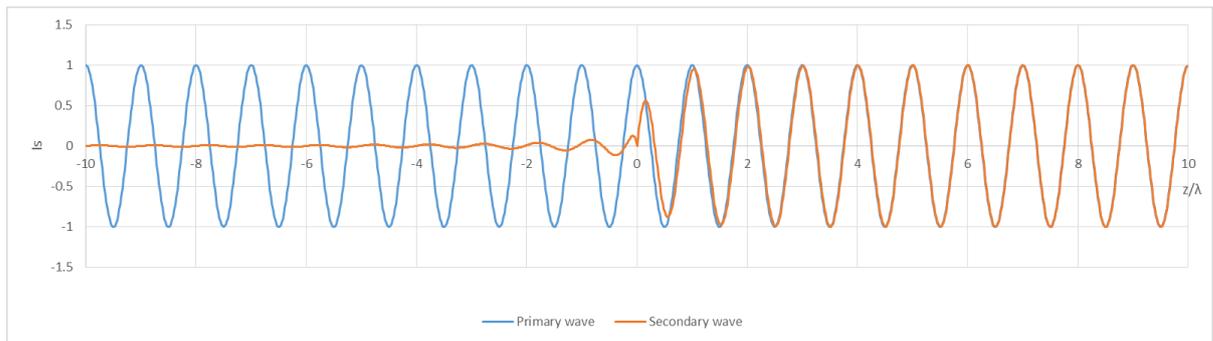

Fig. 6. Wave intensity along Oz in case 5 ($K_0 = \frac{\cos(\alpha)+\cos(\chi)}{2}$).

In case 5 ($K_0 = \frac{\cos(\alpha)+\cos(\chi)}{2}$) the waves coincide as closely as in the previous cases, except the region of $z < 0$, where the resulting wave intensity becomes negligible. The discrepancy between the primary wave and the resulting wave in forward direction is something average between the cases shown above (not shown here). This behavior looks strange with regard of symmetry, but physically reasonable, if to focus on diffraction, rather than reflection.

Fig.7 shows that the variance in magnitudes in all cases is negligible, when the point of observation (P) is located far from the plane of secondary sources.

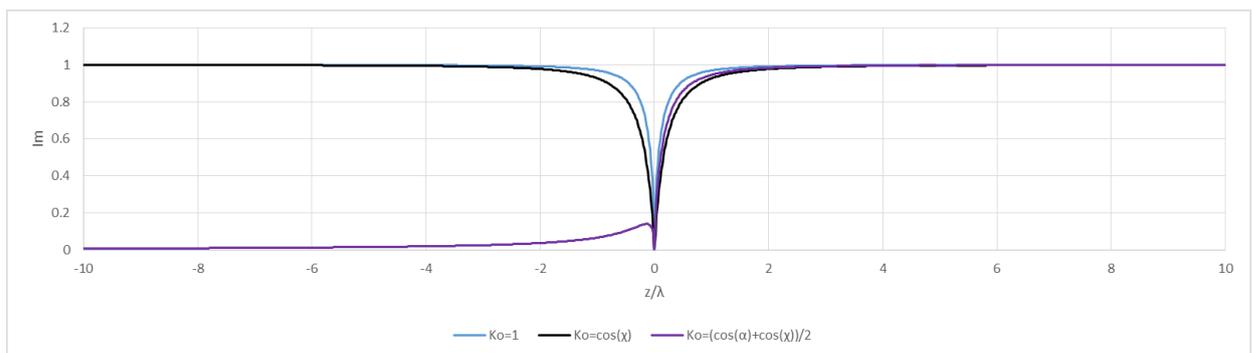

Fig. 7. Magnitudes of the secondary waves in case of plane primary wave.

In the second experiment, secondary waves are located on the same plane, but the intensity of the resulting waves is calculated in point P, positioned on a spherical wave front, as shown in fig.8. The intensity of the primary wave at dS can be calculated here as $I_i/r$, where r is the distance from the source to dS.

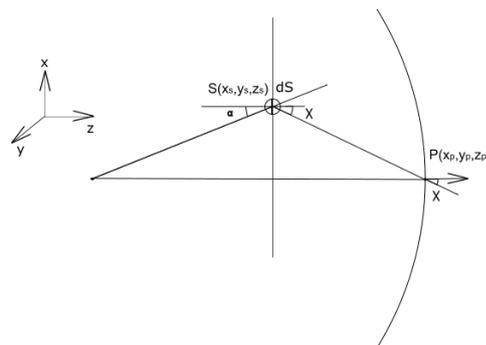

Fig. 8. Simulation of the spherical wave.

The results of the simulation can be seen in fig.9. The waves coincide in phases, thus only magnitude distribution is shown. It could be noted that all five options of $K_0$ produce negligible difference in magnitude in the area of positive z. In the opposite direction, the situation is different: $K_0 = \cos(\chi) \cdot \cos(\alpha)$ (case 4) provides the best approximation, case 3 also can be used (some fringes can be reduced, when the distance between the primary source and the plane with secondary source is increased), while isotropic secondary sources (case 1) give significant fringes. Case 5 has negligible intensity, when $z < 0$.

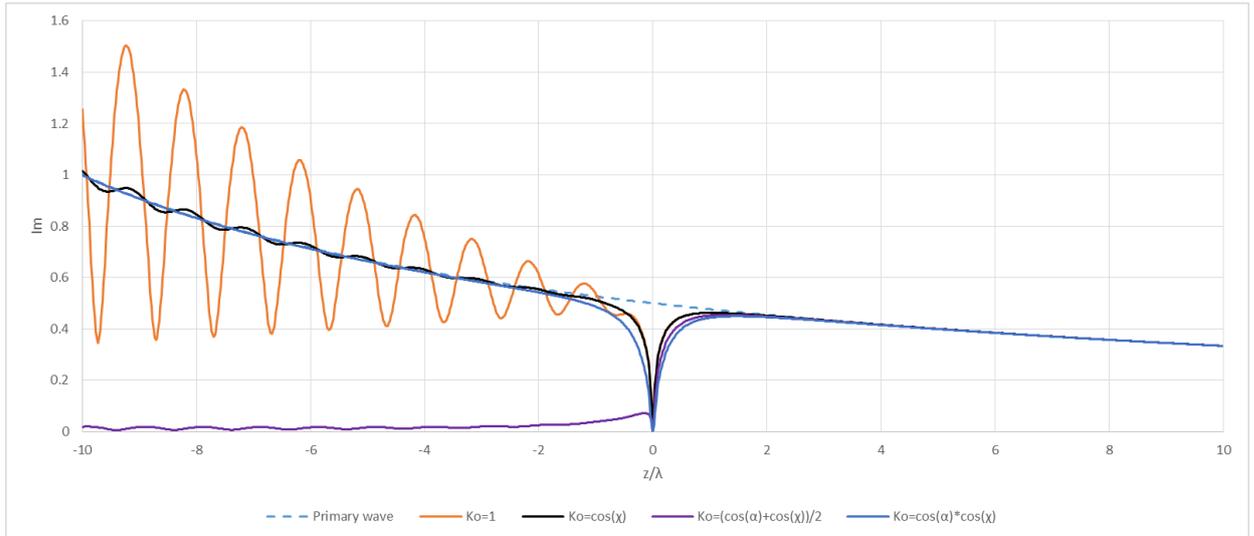

a) Cases 1,3,4,5

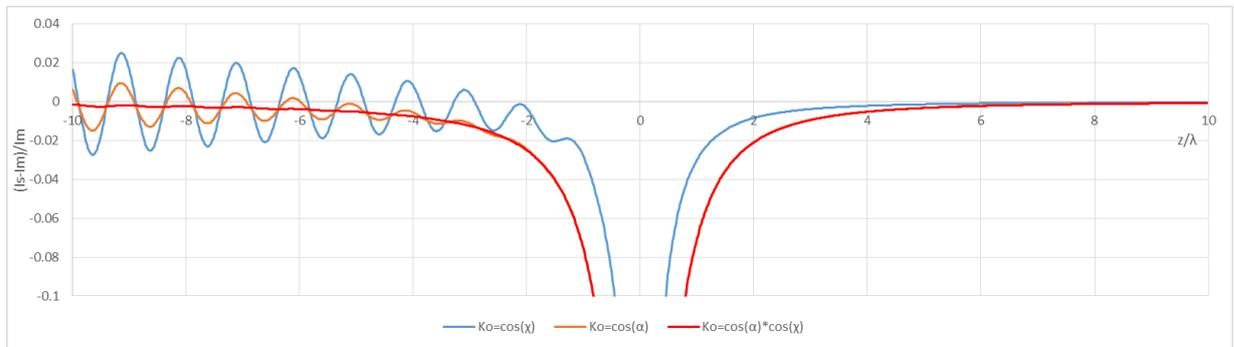

b) Cases 2,3,4 (magnified)

Fig. 9. Magnitudes of the secondary waves in case of the spherical primary wave.

The results of experiments show that in order to obtain the reliable and accurate results in numerical computations, it is possible to substitute plane or spherical waves by a set of secondary sources in accordance with the principle, mentioned above. The choice of an incident factor (obliquity factor) does not play any significant role, when we are talking about distances more than few wavelengths in the direction of the wave propagation. In case of backward direction, $K_0 = \cos(\alpha)$ or $K_0 = \cos(\chi) \cdot \cos(\alpha)$ are appropriate, while $K_0 = \frac{\cos(\alpha)+\cos(\chi)}{2}$ can be used, when we are dealing with the wave with negligible intensity in backward direction.

### 3. Diffraction by a disk

The task is to calculate $U(P)$ as a result of the interference of an incident wave and secondary waves, produced by sources on the surface of the obstacle. This experiment with a similar approach is discussed in [9]. The idea of the experiment is explained in fig. 10. Here the plane wave meets an obstacle (circular disk) normally. Only the wave, scattered in forward direction, contribute into resulting diffraction screen under the disk.

In a physical experiment it is reasonable to use a CMOS or CCD matrix to obtain a diffraction pattern. To make the results of simulation relevant, it is necessary to calculate power, absorbed by the matrix or reflected beyond the matrix. Typically in particle counters the size of the matrix is sufficient to gather most part of scattered light. It is assumed that only the vertical component ($U_z$) contributes to this diffraction pattern. Thus the results of the simulation will be displayed as a function $F(y) = U_z(y)^2$. It is clear that the pattern will be symmetrical due to the circular form of the disk. In this experiment the wave with $\lambda = 0.6 \mu m$ was used.

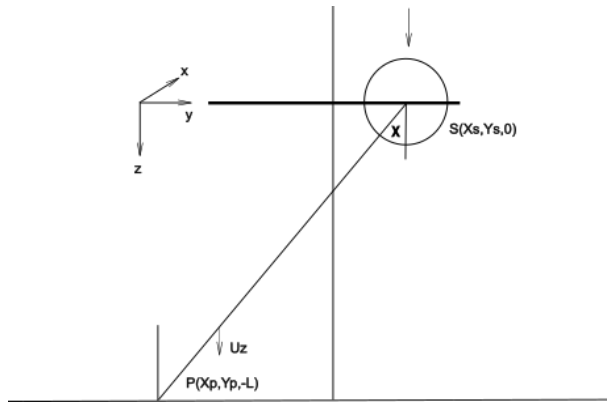

Fig. 10. Simulation of the diffraction of the plane wave by a disk.

Fig. 11 shows, how diffraction pattern depends on the distance from the disk ($60\lambda$ diameter) to the screen. Due to the symmetry, only one half of cross-section is shown (y varies from the centre of the pattern to its edge).

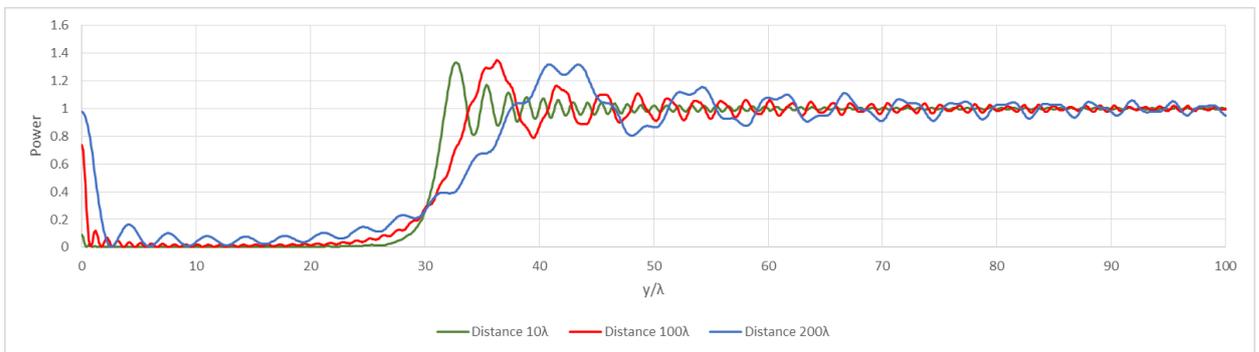

Fig. 11. Diffraction pattern from the disk at various distances to screen (plane wave).

It could be seen that in all cases some Poisson (Arago) spot is present, however its brightness is less, when the disk is located close to screen. It is also seen that brightness oscillations at the central area increase at big distances.

In the experiment with spherical waves the same approach is used, but the wave is irradiated by a source, located at $L$ above the screen. Fig. 12 shows the results of simulations with the same disk, but various values of L. Power is given normalized to make the diagram comparable to fig. 11.

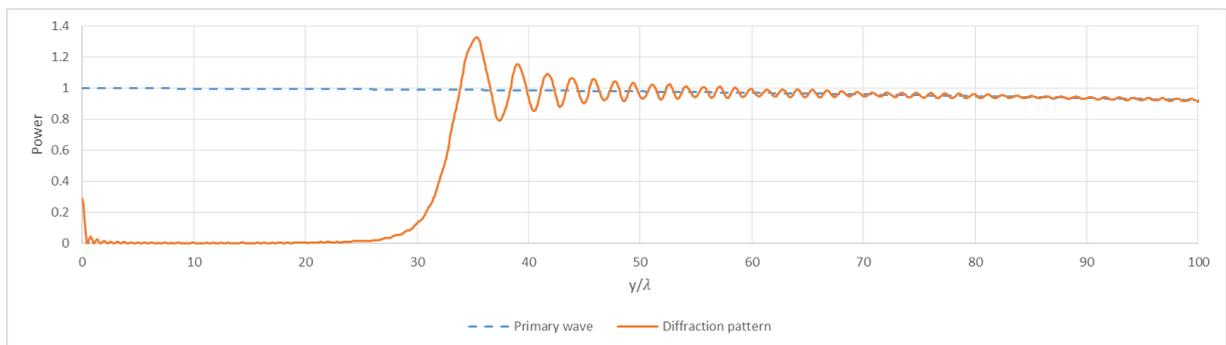

a) Distance $500\lambda$ from disk to the primary source

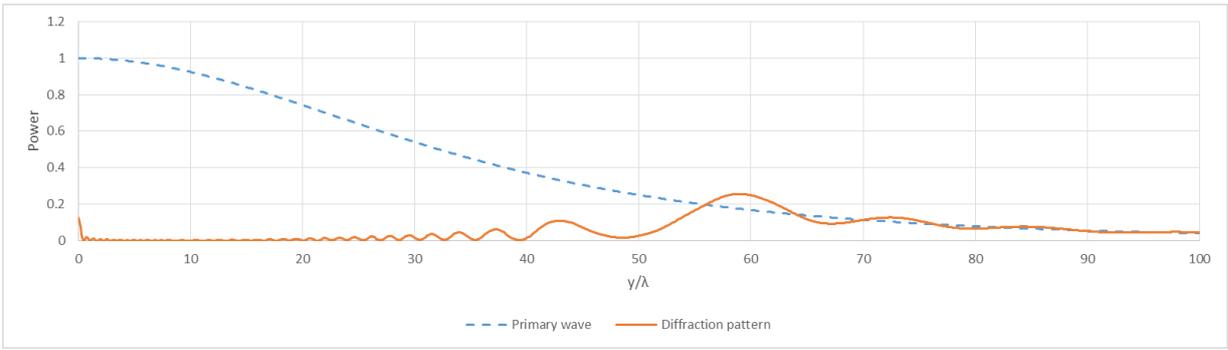

b)  Distance $50\lambda$ from disk to source

Fig. 12. Diffraction patterns of a $60\lambda$ disk in case of a spherical wave.

In experiments with diffraction by a disk, it was noted that obliquity factors play negligible role, except only the area of Poisson spot. This phenomenon can be easily explained by two factors: brightness of the Poisson spot mainly depends on the intensity of secondary waves, radiated at significant angles and we register only the power of the component, normal to screen. Fig.13 shows this area of Poisson spot, where the difference can be seen. Noteworthy that the case $K_0 = \frac{\cos(\alpha)+\cos(\chi)}{2}$ (case 5) gives the same curve, as the one with $K_0 = \cos(\alpha) \cdot \cos(\chi)$ (case 4).

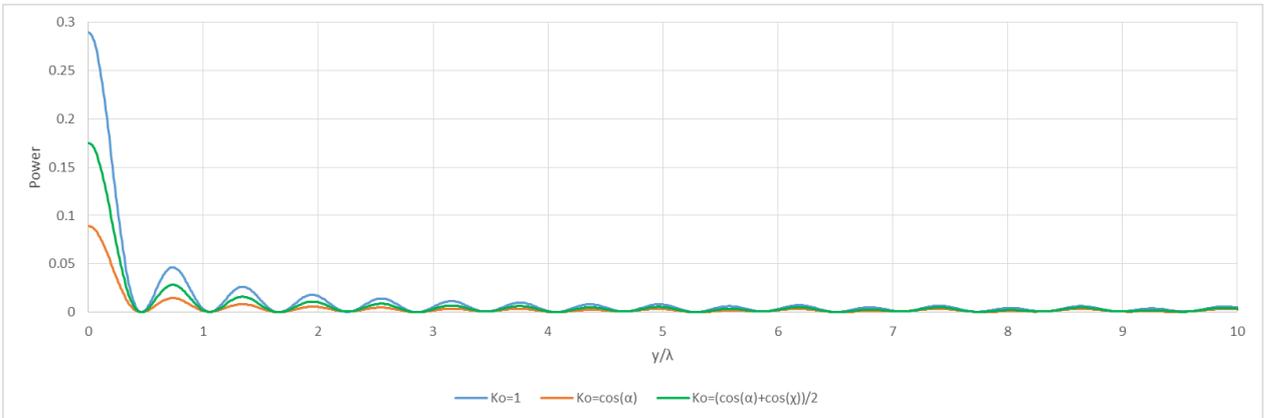

Fig. 13. Poisson spot area of diffraction patterns.

### 4. Diffraction by a sphere

The most studied case is Mie scattering, described in details, for example, in [10]. Mie theory is built upon Maxwell's equations and the scattering process depends on the material of a sphere. Huygens-Fresnel approach in the form, described here, cannot be directly applied to transparent objects. It is essential to suppose that we can use it in cases, when the object absorbs or reflects all the incident power. From the physical point of view, it depends on refractive index, mainly imaginary part of it. Some discussions on various problems of this kind can be found in [11,12].

Mie theory gives us angular patterns in far field, rather than diffraction patterns, needed in some particular applications, as particle counters. Using well-known BHMIE program [10] it is possible to compute angular scattering for various cases. Some results are shown in fig. 14. These results are obtained for a sphere of $24\mu m$ in diameter with refractive index $N = n + jm$, where $n = 0.57$ (gold [11]) for the wavelength $0.55\mu m$. Several options for $m$ were investigated. It could be seen that the scattering process depends on $m$ mainly in the area of sidelobes, where the magnitude (normalized to the intensity of scattering forward) is 20-30 dB less, that the intensity of the scattering in forward direction. These sidelobes become less for big spheres. Cases $m = -2.45$ and $m = -24.5$ display little difference. BHMIE provides far field results, which are not applicable in some practical situations (near field), however it can be used to verify calculations results of the proposed computational algorithm.

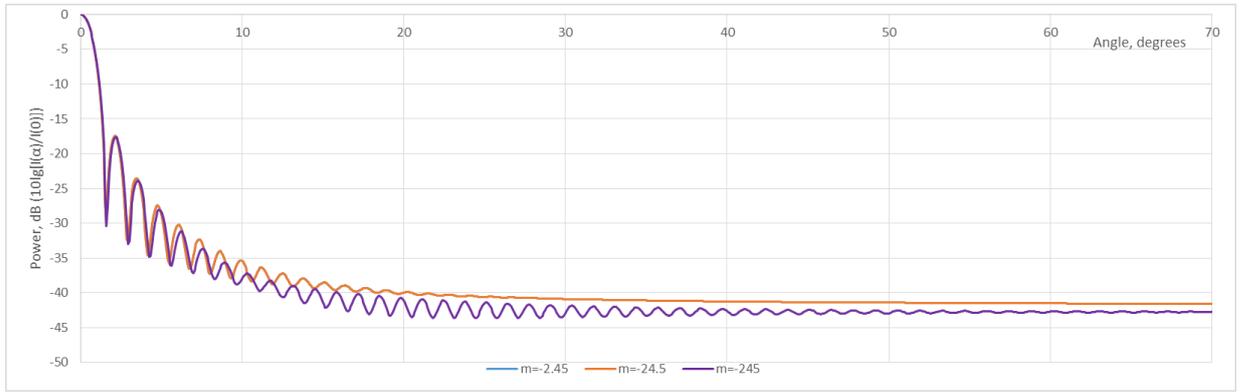

Fig. 14. Angular scattering by a sphere (Mie theory).

A similar result can be obtained, using described algorithm. In this experiment only the surface, visible from P, contributes to the resulting intensity. All the calculations are performed in accordance to fig.15.

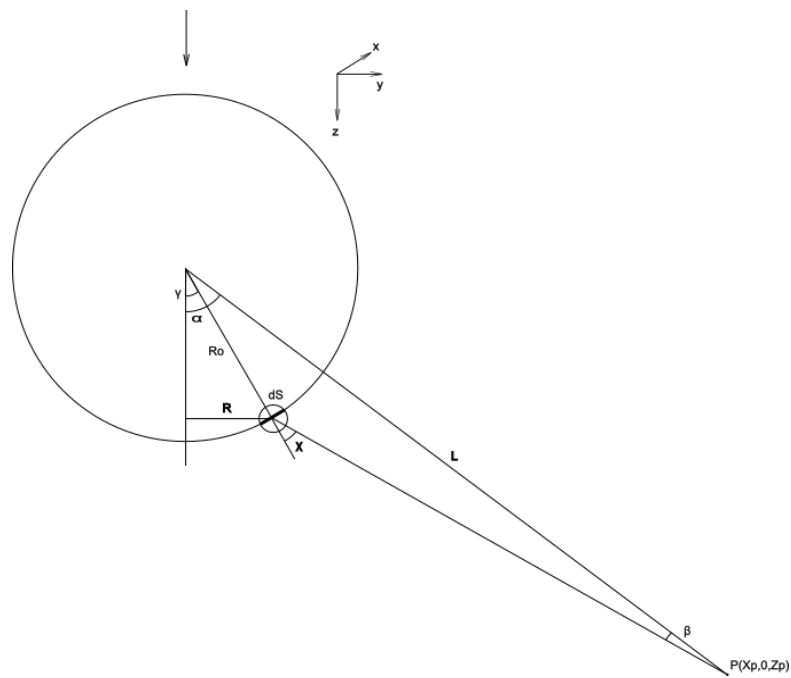

Fig. 15. Simulation of the diffraction by a sphere.

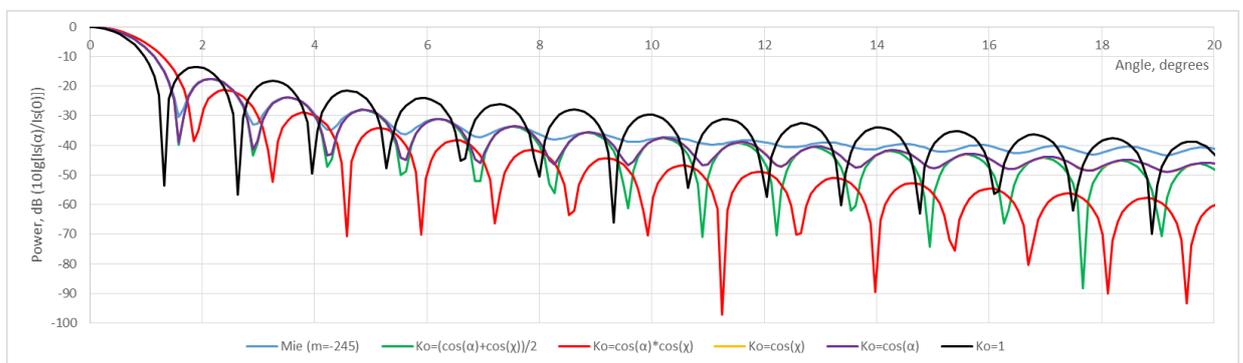

Fig. 16. Angular scattering by a sphere (diameter $24 \mu m$).

Fig. 16 shows that cases 2,3,5 are close to Mie results (with m=-245.0) and case 2 is almost identical to case 3, so the difference between them cannot be seen. In general, power of the diffracted wave, calculated with Huygens-Fresnel algorithm is much less than the one, calculated by BHMIE outside the main lobe of the diagram. This inference is also confirmed in [10,13]. Fig. 17 displays almost the same results for a smaller sphere.

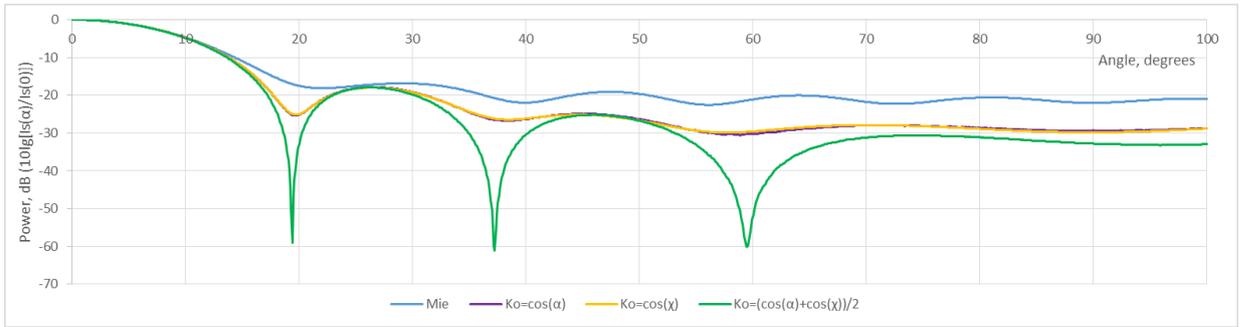

Fig. 17. Huygens-Fresnel angular scattering by a sphere (diameter $2\mu m$).

Huygens-Fresnel simulation can be easily used to calculate angular scattering pattern. Fig. 18 displays some cases. It is seen that scattering profile resembles Mie results only at the distances exceeding several hundreds of wavelengths.

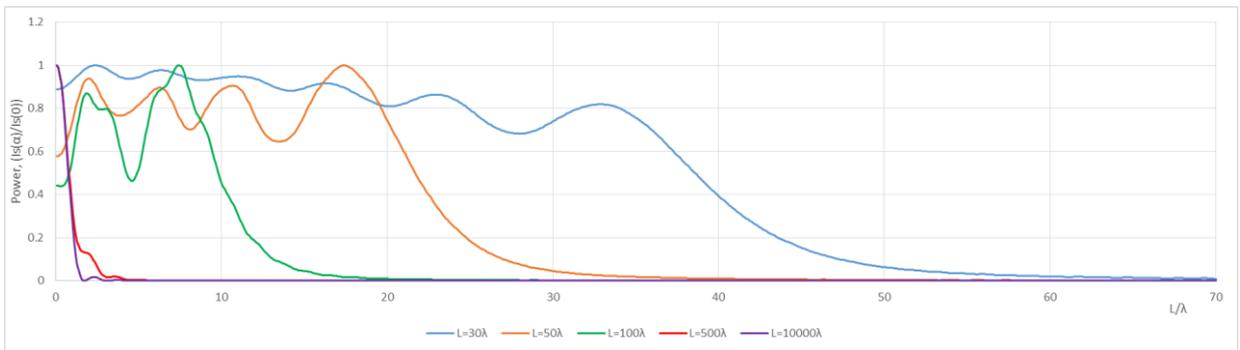

Fig. 18. Dependence of the scattering on the distance to the observation point (sphere $24\mu m$).

According to [10], big absorbing spheres scatter as the disks of the same radii. This comparison for the spheres of 2 $\mu m$ and 24 $\mu m$ in diameter is given in fig. 19.

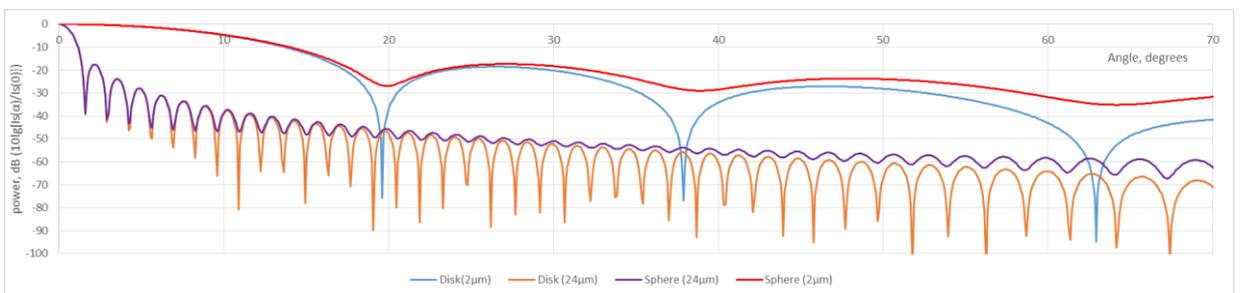

Fig. 19. Angular scattering by a disk and a sphere.

Fig. 20 shows the results of comparison of the simulation of diffraction by disks and spheres. The patterns look similar, however it can be seen that the difference become greater in the near field.

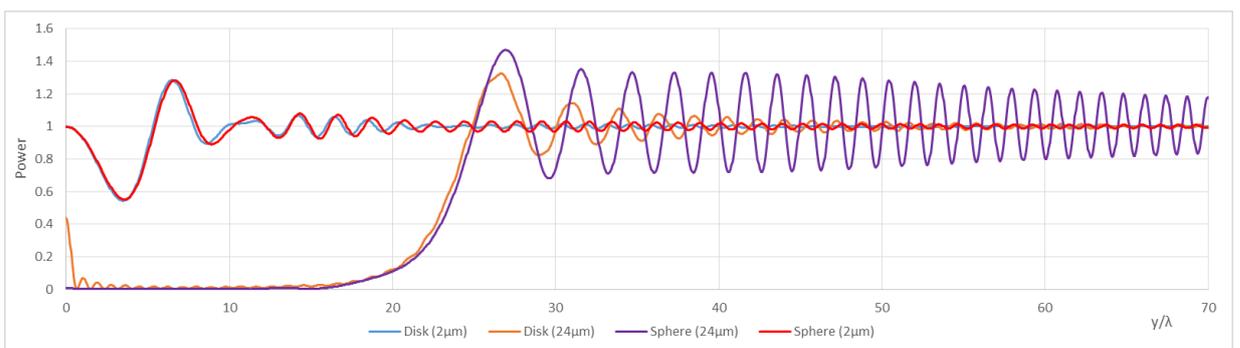

Fig. 20. Diffraction patterns of a disk and a sphere (distance 30 $\lambda$ to screen).

## 5. Summary

The numerical experiments, shown above, show that the algorithm of direct simulation, based on Huygens-Fresnel-Kirchoff principle, can be used in many cases along with more sophisticated approaches. Using this simulation, we have to assume that we are dealing with absorbing or reflecting obstacles without any information about the substance, of which they consist. It can be applied to the objects of any form in both far and near fields; however the paper did not investigate very small obstacles ($d < \lambda$). It is interesting that Fresnel's idea about the directivity of secondary sources very likely does work in many practical cases, while the obliquity factor does not contribute much.

This paper was inspired by some practical researches in the area of industrial particle counters. In this application area the main objects are opaque and relatively large ($d > \lambda$). The simplicity of the proposed approach allows extrapolating it to the more specific cases of non-coherent and non-monochromatic sources with minimal problems.